\documentclass[conference,compsoc]{IEEEtran}
\IEEEoverridecommandlockouts
\usepackage{cite}
\usepackage{amsmath,amssymb,amsfonts}
\usepackage{algorithmic}
\usepackage{graphicx}
\usepackage{textcomp}
\usepackage{xcolor}
\usepackage{pifont}
\usepackage{listings}
\usepackage{tabularx}
\usepackage{booktabs}
\usepackage{ulem}
\usepackage{lipsum}
\usepackage{enumitem}

\usepackage{tikz}
\usepackage{pgfplots}
\usepackage{xcolor}
\pgfplotsset{compat=1.18}
\usepackage{pgfplotstable} 

\usepackage{tikz}
\newcommand*\circled[1]{\tikz[baseline=(char.base)]{
    \node[shape=circle,draw,inner sep=1pt] (char) {#1};}}

\usepackage{hyperref}
\usepackage[%
  square,        
  comma,         
  numbers,       
  sort&compress 
]{natbib}

\usepackage{url}
\usepackage{ulem}

\newif\ifdraft
\drafttrue
\ifdraft

  \usepackage{soul}
  \usepackage{etoolbox} 
  \usepackage{tcolorbox}
  \usepackage{xcolor}
\newtcolorbox{msnotes}{
  colback=white,
  colframe=blue,
  fontupper=\color{blue},
  boxrule=0.5pt,
  arc=0mm,
  boxsep=1pt,
  left=1pt,
  right=1pt,
  top=1pt,
  bottom=1pt,
  before upper={\textcolor{blue}{MS:\\ }}, 
}
\newcommand{\msnote}[1]{ {\textcolor{blue} { ***MS: #1 }}}

\newcommand{\bsnote}[1]{ {\textcolor{red} { ***BS: #1 }}}
\newcommand{\red}[1]{ {\textcolor{red} {\em #1 }}}
\else
\newcommand{\msnote}[1]{}
\newcommand{\bsnote}[1]{}
\newcommand{\lfnote}[1]{}
\newcommand{\red}[1]{}
\fi

\def\BibTeX{{\rm B\kern-.05em{\sc i\kern-.025em b}\kern-.08em
    T\kern-.1667em\lower.7ex\hbox{E}\kern-.125emX}}
\begin{document}

\title{Influence of Logging Frameworks on Bind9}
\author{\IEEEauthorblockN{Max Schrötter}
\IEEEauthorblockA{\textit{Department of Computer Science} \\
\textit{University of Potsdam}\\
Potsdam, Germany \\
schroetter@cs.uni-potsdam.de}
\and
\IEEEauthorblockN{Hannes Signer}
\IEEEauthorblockA{\textit{Department of Computer Science} \\
\textit{University of Potsdam}\\
Potsdam, Germany \\
hannes.signer@uni-potsdam.de}
\and
\IEEEauthorblockN{Bettina Schnor}
\IEEEauthorblockA{\textit{Department of Computer Science} \\
\textit{University of Potsdam}\\
Potsdam, Germany \\
schnor@cs.uni-potsdam.de}}

\maketitle

\begin{abstract}

Host-based Intrusion Prevention Systems (IPS) rely on application logs to
detect and block malicious activity. However, on modern high-speed networks the
logging subsystem itself becomes a bottleneck: an attacker can hide traces
simply by generating enough traffic to overwhelm the application's log
pipeline, dropping crucial traces. In this work, we show that widely deployed
setups such as Fail2Ban monitoring BIND9 can be defeated with less than 65~Mbps
of DNS traffic. Further, we show that when replacing core components of the IPS
architecture with their higher-performance equivalent, iptables with eBPF and
regex matching with Hyperscan, the logging backends themselves become the
bottleneck. Therefore, we present FIPS, a new IPC designed for high-performance logging that
bypasses the kernel and reduces copying of the log messages to a minimum. FIPS uses
per-thread lock free shared memory ring buffers, supporting multiple
independent consumers reading the same log stream at their own pace. FIPS
offers both a native API and a drop-in replacement for the syslog interface.
Our evaluation with BIND~9 shows that FIPS introduces almost no overhead
compared to disabled logging, logs more requests than any other evaluated
framework, and enables the IPS to ban malicious clients {\boldmath $2.5\times$} faster than
with file logging while sustaining {\boldmath $2^{16}$} attacking clients at one million
requests per second.
\end{abstract}

\begin{IEEEkeywords}
Intrusion Prevention, 100GbE, IPC, Logging
\end{IEEEkeywords}

\section{Introduction}

\begin{figure*}[ht]
  \centering
  \includegraphics[width=0.9\textwidth]{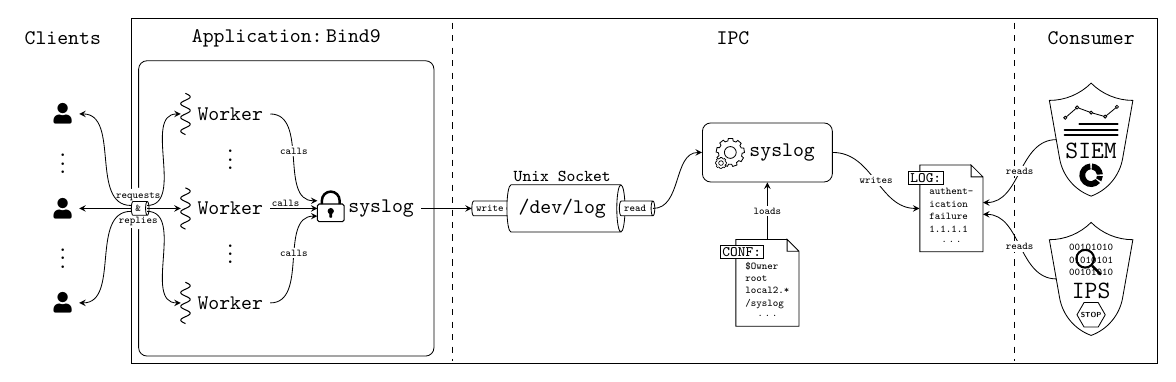}
  \caption{Overview of the classic IPS workflow}
  \label{fig:overview}

\end{figure*}

The detection and prevention of intrusions is a key part of securing computer
systems. Depending on the location and function of the intrusion detection
system it uses different data sources. Network intrusion detection systems use
network traffic to detect attacks, while
the analysis of system and application logs is the key component of host intrusion detection. Applications usually
write text-based log files that are analyzed by log collectors (like
logstash, beats, fluentd, etc.) and then forwarded to SIEMs and IDSs.
These log collectors parse the log files from  strings into
fields and store them in databases in a unified and machine-readable format for later analysis.
Figure \ref{fig:overview} shows this workflow
with BIND9 as an example application and syslog as logging backend.
While the correct parsing of logs is its own research topic
\cite{log-parsing-evaluation}, a lot of effort could be circumvented by
consequently logging in a user-definable binary or structured format.


Application logging is a core component of every user-space application.
The most widespread logging options are file logging, utilizing a variation
of the write syscall and syslog, utilizing either a libc syslog
function\cite{apache,bind9} or
reimplementing a variation\cite{nginx}.
While syslog is, strictly speaking, a protocol (RFC 5424~\cite{rfc5424}), most
current logging daemons (e.g. rsyslogd or systemd-journald) create the
\texttt{/dev/log} Unix socket to receive new local log messages. While most
applications will by default write messages to this Unix socket, the transport
via UDP and TCP is also widely supported. Local logging
daemons process new logs and might also forward them to dedicated logging servers.
A variation of file logging which is almost universally supported is the logging to
the stdout or stderr filedescriptor. Current Init-Systems like systemd read
those messages and forward them to the local logging daemon.

Logs themselves are a primary target for attackers to hide their
attacks~\cite{carbon-black}. This can be accomplished in multiple ways:
modifying the logs after a system has been compromised or overwhelming a
component in the logging pipeline. Securing logs against tampering after
a system has been compromised has been thoroughly covered in the
literature~\cite{hardlog, omnilog, nitro, kennyloggings, sgxlog, tpm-log}.
This work focuses on how easily the application or the used IPC
can be overwhelmed to hide traces of an attacker.

Sommese et al.~\cite{ddos_dns_infrastructure} showed that DNS infrastructure is a
common target for DDoS attacks. They also showed that affected DNS hosting
providers are small to medium-sized. While complete failure in DNS name
resolution is rare (1\%) the  attacks increase the RTT up to 348
times.

In another scenario, an attacker might generate a high volume of DNS logs to hide
the execution of malware. Malware uses DNS and Domain Generation Algorithms
to resolve addresses of Control and Command Servers.
An Intrusion Prevention System (IPS) and the logging framework must be able to
cope with high volumes of logs and employ mechanisms to reduce log volume
like rate limiting to mitigate these attacks.

In this work, we focus on the full stack, from the application producing the
logs, to the Inter-Process Communication (IPC) used to transport them,
and to the log collectors.
Our contributions are:
\begin{enumerate}
\item We show how easily a widely used IPS like Fail2Ban can be
  overwhelmed.  (see Section~\ref{sec:limitations})
\item We evaluate the performance impact of the logging framework on
  the performance of BIND9.  (see Section~\ref{sec:limitations})
\item Our analysis shows that the logging backends themselves become
  the bottleneck.  Therefore, we propose a new IPC called FIPS IPC
  designed for high-performance logging that bypasses the kernel.
  FIPS uses per-thread lock-free shared-memory ring buffers,
  supporting multiple independent consumers reading the same log
  stream at their own pace. The memory containing the ring buffers is
  directly shared between the logging application and the consumers,
  reducing memory copies of the log messages.  (see
  Section~\ref{sec:FIPS})
\item We evaluate the new IPC against state-of-the-art logging
  frameworks and show its performance benefit both on the performance
  of the IPS and on the application (BIND9).

  Fail2Ban is written in Python and, because of the interpreted
  language and slow regex pattern matching, is not designed for
  high-performance. In the evaluation we, therefore, replace it by a
  simplified C-based implementation (simple-Fail2Ban). Core components
  were replaced with their more efficient equivalents, iptables
  with eBPF and regex matching with Hyperscan~\cite{hyperscan}.

We analyze how well our proposed FIPS IPC with BIND9 and the improved
simple-Fail2Ban is able to cope with DoS attacks up to 10 million
packets per second. (see Section~\ref{sec:evaluation})

\item
  As DoS mitigation strategy the DoS
  clients are rate limited. We also analyze the number of logged
  requests to see if traces of answered requests can be lost.
  (see Section~\ref{sec:evaluation})
\end{enumerate}
\section{Limitations of IPS in High-Speed Networks}\label{sec:limitations}



In our first experiment, we use the widely used IPS Fail2Ban~\cite{fail2ban} as
baseline. Fail2Ban uses user-definable regular expressions to detect
patterns in log files and executes a user-definable action if a pattern
is detected, as well as another action after a user-definable time
interval. One of the most widely used configurations is to monitor
authentication tries against sshd or a web server. If a user enters
incorrect credentials more often than a configurable limit within a
configurable time frame, the IP address is banned via
iptables.

Fail2Ban is however not limited to authentication brute
force attacks, but supports also rate limiting.
The advantage of utilizing Fail2Ban for rate limiting is that requests
that exceed the rate limit are blocked at the firewall level, which
frees up resources for the application, for example a DNS server.
BIND9 is configured to log all
DNS queries to a file and Fail2Ban monitors this file for clients that
have exceeded the rate limit.

To test the performance of this system, we set the rate limit to just
3 requests per 60 seconds. This limit is well below the rate for common
DNS clients. Web browsers, for example, often send multiple DNS
requests for a single web page to load. First,  DNS requests for
the IPv4 and IPv6 records for the
requested domain are sent, and additional requests for resources like webfonts,
javascript libraries, and third party tracking \& integrations follow, which
are often hosted under different domains (CDNs). Furthermore, we set a
ban duration for 60 seconds, whereas the default rate limit window of BIND9 is
15 seconds. This configuration gives Fail2Ban the opportunity to ban
clients earlier and longer and  reduces the load on the DNS server.

\begin{figure}
  \begin{center}
    \includegraphics[width=0.45\textwidth]{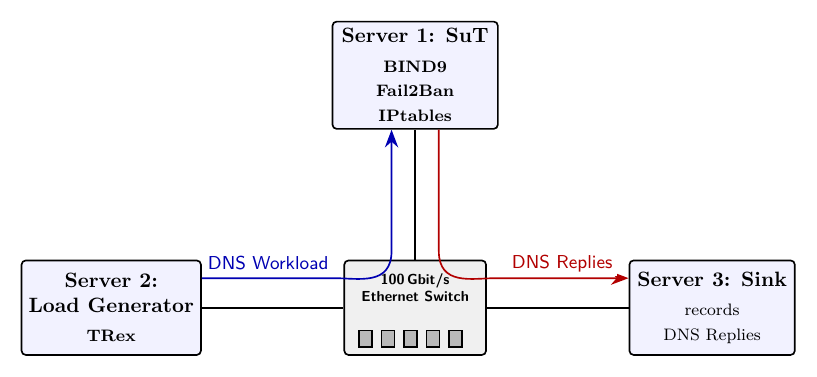}
    \caption{Test Bed}%
    \label{fig:testbed}
  \end{center}
\end{figure}

All measurements are done on servers with an
Intel(R) Xeon(R) Silver 4314 Chip with 16 cores and 128 GB RAM and
ConnectX-6 Dx Cards connected via 100 GBit Ethernet.
DNS requests were generated with Trex~\cite{trex}.
The test bed is shown in Figure~\ref{fig:testbed}.

First, we test this application scenario  under attack and identify the 
  bottleneck.


\subsection{Ban Capabilities under Attack}\label{sec:f2b-performance}

For this experiment, Trex generated traffic from two different source
networks. One network generates legitimate baseload of 50,000 requests per
second. From the second network we simulate an attacker which sends 100,000
requests per second from 255 different IP addresses which is much more
than a DNS server is capable to handle.

\begin{figure}
  \begin{center}
    \includegraphics[width=0.45\textwidth]{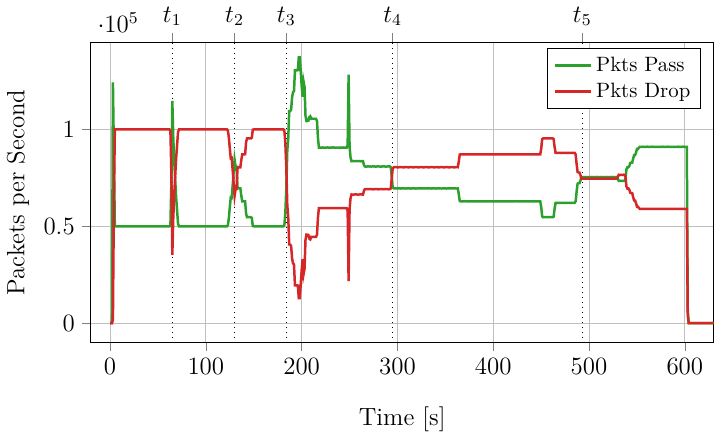}
    \caption{Fail2Ban Performance Investigation with 255 spoofed IP addresses
    and 100000 packets per second from attackers}%
    \label{fig:f2b-baseline}
  \end{center}
\end{figure}
Figure~\ref{fig:f2b-baseline} shows that Fail2Ban is initially able to parse the
BIND9 file logs and ban all attacking clients. After a spike in the \texttt{Pkts
Pass} curve in the first few seconds, the firewall drops all 100,000 requests
per second from the attacking clients. At $t_1$ the 60 second ban time has
elapsed and fail2ban allows the previously banned clients again.
The figure also shows that it takes Fail2Ban longer in the second ($t_2$) and
third ($t_3$) ban
cycle to ban all attacking clients.
After the third ban cycle ($t_3$) Fail2Ban completely fails to keep up with the log
volume and is not able to ban all attacking clients anymore.

\begin{figure}
  \begin{center}
    \includegraphics[width=0.45\textwidth]{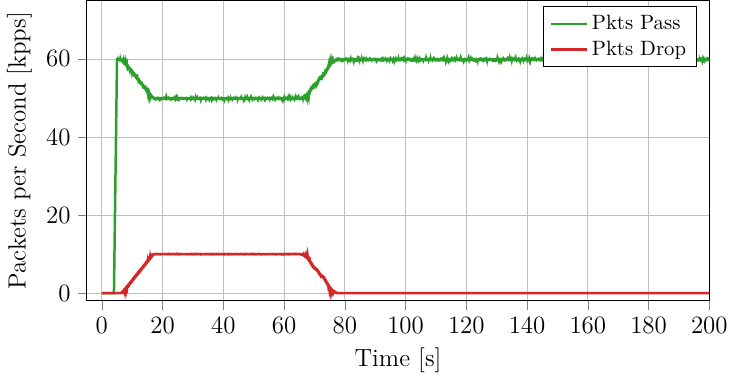}
      \caption{Fail2Ban Performance Investigation with 2047 spoofed IPs and
      10000 packets per seconds from an attacker}%
    \label{fig:f2b-baseline2}
  \end{center}
\end{figure}

If the attacker increases the number of spoofed IP addresses to 2046 this
behavior already occurs at 10,000 packets per second, as can be seen in
Figure~\ref{fig:f2b-baseline2}. The figure shows only one ban cycle (10s-70s) in which
Fail2Ban bans the attack traffic. It also takes Fail2Ban significantly longer to
ban 2047 addresses with a tenth of packets per seconds, than 255 clients. 

Even though 100,000
requests per second sounds like a high volume of traffic, it is actually less
than 65~Mbps with each DNS request being 648 bits long.

Profiling the system under stress showed that 15.1\%\footnote{using
profile-bpfcc to profile user-space and kernel-space} of total CPU time
was spend in the BIND9 logging function \texttt{isc\_log\_doit}. Even
under high load, where BIND9 is not able to reply to all requests, the system is
still 65\%\footnote{measured with mpstat} idle. To determine why
BIND9 is not using more CPU time an off-CPU analysis\footnote{using
offcputime-bpfcc} was performed. It showed
that 80\% of the time, a BIND9 thread could not run because of the lock inside
\texttt{isc\_log\_doit}.
This confirms that logging is a bottleneck, which we investigate further in the next section.
%
%
%
%

\subsection{Impact of Logging}\label{sec:logging-impact}



In this subsection the performance impact of different logging options is
analyzed. BIND9 supports multiple logging options, as many Linux applications,
i.e.\ file logging, syslog, systemd journal. For all logging frameworks BIND9 supports a
configurable log format and different log levels. BIND9 is multi-threaded, but the
logging function is protected by the global ``\texttt{isc\_\_lctx->lock}''
mutex. This is done for non thread-safe logging
functions and the use of a single global log buffer. 

For file logging, BIND9 uses fprintf to write to the open file stream.
For syslog, BIND9 calls the syslog function from the libc.  There, the
syslog messages are sent via the syslog (/dev/log) socket to a syslog
daemon which handles the message further.  The syslog daemon
was configured to write the logs to a file.

For the systemd journal, messages are printed to stdout and captured by systemd
which writes the messages to memory, and only if configured writes it
to persistent storage.  Unlike syslog,
the systemd journal was configured to use volatile storage.  

\begin{figure}[t]
  \centering
  \includegraphics[width=0.5\textwidth]{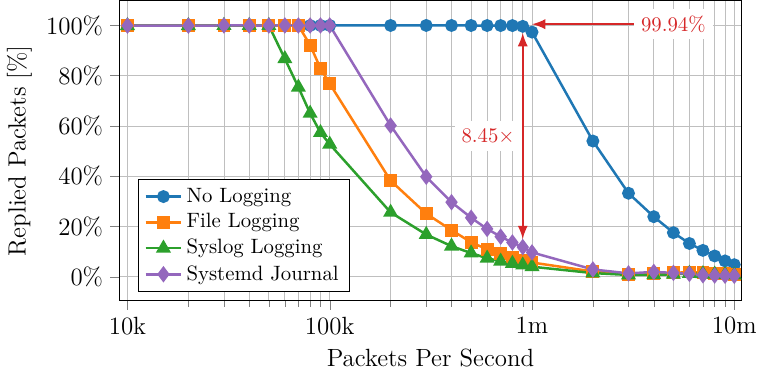}
  \caption{Logging Performance Comparison of BIND9}
  \label{fig:static-ipc-intro}
\end{figure}


In this experiment, TREX generates only benign traffic from 255 IP addresses,
  starting from 10,000 up to 10 million packets per second.
The following figures present the median of 3 measurements.
Figure~\ref{fig:static-ipc-intro} shows the performance impact of the different logging
frameworks on BIND's performance. BIND9 with no logging configured replies to 99.94\% of requests when
900,000 requests per second are being sent. If BIND9 is using logging via
journald the reply rate drops to
11.79\%. This is an 8.45x performance drop. The results for syslog
and file logging are worse with only 4.64\% and 6.53\% of requests being
answered respectively.
The journald performance shows the advantages of in-memory logging over
file-based logging. While the syslog performance shows the disadvantages of the
extra overhead of the Unix socket and logging daemon before writing to disk.

However, all available logging options show a significant performance drop compared to no
logging. This shows that logging is one of the bottlenecks for Intrusion
Prevention Systems.

\section{Related Work}\label{sec:related-work}

There has been little change in logging functions for widely used
applications like Apache httpd~\cite{apache}, nginx~\cite{nginx}, BIND9~\cite{bind9}
and many more. However, there have been additions in logging formats for example qlog~\cite{qlog} the
``Structured Logging for Network Protocols'' from the IETF Quic working group.
Also domain specific logging formats and transport have been introduced, for example DNSTAP~\cite{dnstap} for logging
DNS events relying on Googles Protocol Buffers and forwarding these events over
ring buffers to a stream socket.

Even though there has been little development in existing applications the topic
is nonetheless heavily researched. Jaein Jeong~\cite{jeong} has identified the
need for high performance logging for GNU/Linux Applications in 2013 and
proposed a shared-memory logging framework. Applications enqueue log messages in a memory
mapped file from which the proposed mqlogd dequeues them and forwards them via
named pipes to a designated output utility like syslogd. For their evaluation a
flash storage logger was used as output utility and different locks for
synchronization. The pthread mutex lock implementation outperformed syslogd by a
factor of 4.0.

Although Domain Name System (DNS) query logs are a critical resource for
detecting malware deployments within a network, enabling text-based logging on
high-traffic authoritative DNS servers, such as BIND9, significantly degrades
their query-resolution throughput. To resolve this, the DNSTAP protocol
was introduced as a high-speed binary logging architecture~\cite{dnstap}. DNSTAP encapsulates
raw DNS events using Google Protocol Buffers~\cite{protobuf} and transports them asynchronously via Unix domain
sockets using the \texttt{fstrm} library~\cite{fstrm}.
Its reliance on Unix domain sockets inherently establishes a 1-to-1 stream.
If multiple consumers require access to the logs, an external software multiplexer must be
introduced. The reliance on sockets also include the overhead of a context
switch when writing or reading to and from them.

A distinct branch of research focuses on the
cryptographic integrity of system logs to prevent manipulation by a
privileged adversary~\cite{kennyloggings,nitro,omnilog,hardlog}.
These have been primarily focused on kernel audit logs.  



In the
following, we will discuss advances in High-Performance Queues, which
could be utilized for logging, and Malware Detection based on Query
Logs.
The pursuit of low latency IPC has driven extensive research into
user-space, shared-memory data structures that bypass the Linux
kernel.  There is the ``linuxrb'' ring buffer in the Linux
kernel~\cite{linuxrb}, ``dpdkrb'' as part of DPDK~\cite{dpdk},
``scqd'', a scalable and memory efficient lock-free
queue~\cite{nikolaev} and the queues of the folly~\cite{folly} and
boost~\cite{boost} libraries. However, as Wang et al.~\cite{bbq}
showed they all suffer from consumer contention, producer contention,
delayed tail updates or cache interference between enqueues and
dequeues.  Further, all these approaches support only the classical
producer-consumer. However, for logging it is crucial to support
redundant readers that read and analyse the same content.

Redundant readers are supported by Disruptor~ \cite{disruptor}
and BBQ~\cite{bbq}.
The Disruptor framework was introduced
cope with the need for high performance IPC in financial exchanges by LMAX. It relies on a
self coined term called ``mechanical sympathy''. Instead of expensive
locks or ``Compare and Swap'' (CaS) instructions, the disruptor tries
to mostly rely on memory barriers.  This works for a single
producer. However, if multiple threads are producing items the
disruptor utilitizes CaS instructions so the threads get the next free
slot in the ring buffer in a race-free manner. Furthermore, by
preallocating the ring buffer used for the queue, latency hits for
reallocating new elements can be circumvented and the overhead of Java's
garbage collector, since the pre-allocated ring buffer is
immortal. Chronicles RingZero~\cite{ringzero} outperforms the
disruptor framework by reducing the feature set to a single producer
and a single consumer avoiding the synchronization overhead.

BBQ stands for Block-Based Bounded Queue where the
ring buffer is split into multiple blocks. By spliting the ring buffer
into blocks the four pointers for queue management
(allocated, committed, reserved and
consumed) may point to cells in different blocks. If these pointers
point to cells in different blocks, the consumer just needs to check
if the committed pointer is still in a different block, which has no
cache line interference with writes to the offset of the committed
pointer inside its block.  Only once the producer moves the committed
pointer to a different block the cache lines of the consumer become
invalid.
However, multiple consumers still contend on the pointer \texttt{allocated} and
\texttt{committed}. However, Wang et al. use, what the authors view as in
principle cheaper FAA (Fetch-and-Add)
instructions instead of CaS instructions, because writers can only enter a block if the block
has been previously consumed. Schweizer et al.~\cite{schweizer} have shown that
all tested atomics have comparable latency and bandwidth.
Therefore, we propose the use of multiple ring buffers (one ring buffer per
producer) instead of a single ring buffer with writer contention.

%
%
%
%

\section{Design of the new IPC for High-Performance Logging}\label{sec:FIPS}


To overcome the limitations of existing logging frameworks, this work introduces
a new IPC designed for high-performance logging, while keeping the need of IPS
and SIEMs in mind. The design goals of the new IPC are:
\begin{enumerate}[label=(\alph*)]
  \item Increase logging performance to cope with DoS attacks in
    high-speed performance networks by using lock-free ring buffers (results
    shown in
    Section~\ref{sec:impact-ips})
  \item Support a variant of the multiple producer multiple consumer model
    where  multiple consumers may read independently all available log messages.
    This allows multiple applications to analyse the same log message.
    (see Section~\ref{sec:sync})
  \item Provide a simple interface for applications to use the new IPC
    (see Section~\ref{subsec:Writer} and \ref{subsec:Reader})
  \item\label{itm:plugandplay} Provide a plug-and-play solution for existing applications (see Section~\ref{subsec:Writer})
\end{enumerate}

As shown in Figure~\ref{fig:overview} before, multiple applications may be interested in
the logs generated by the producer application. These can be different IPS systems or
SIEMs or log consumers that forward the logs to a central logging server.
Each of these applications should be able to read all log messages at their own pace.
File-based logging supports this use-case by nature as multiple applications
can open the same log file and read it at their own pace. 

\begin{figure}[t]
  \centering
  \includegraphics[width=0.5\textwidth]{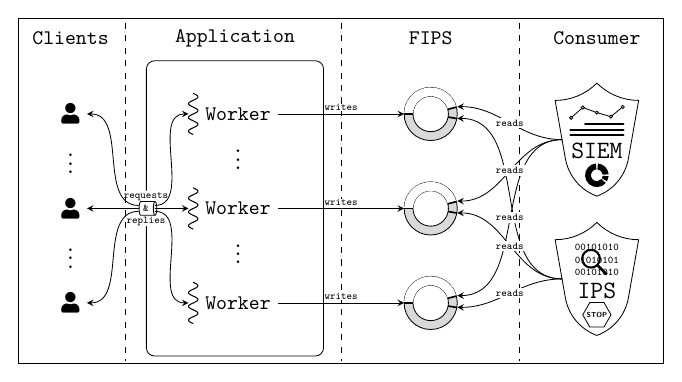}
  \caption{Overview of the new fast IPS architecture}%
  \label{fig:overview-fips}
\end{figure}

The architecture of the new FIPS IPC, an IPC for fast intrusion prevention,  is shown in Figure~\ref{fig:overview-fips}.
To increase logging performance each producer application thread has a
dedicated lock-free ring buffer to write its log messages to.
The new IPC  supports multiple consumers with each having multiple
threads reading from the ring buffers.
In the following, we present the new IPC in detail

\subsection{FIPS IPC Synchronization}\label{sec:sync}

\begin{figure}
  \centering
  \includegraphics[width=0.5\textwidth]{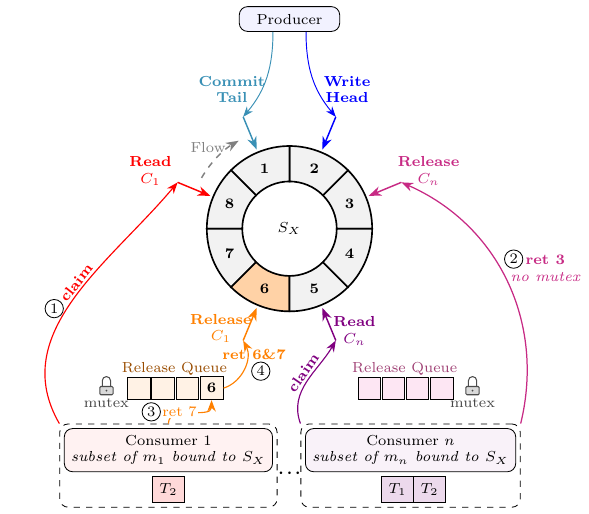}
  \caption{Ring buffer architecture}%
  \label{fig:architecture}
\end{figure}


As previously shown in Section~\ref{sec:f2b-performance}, writing of query logs can limit performance drastically.
This behavior is not limited to BIND9. For example,  nginx  spends 50\%\footnote{measured with
profile-bpfcc to profile user-space and kernel-space} of its CPU time
writing query logs when serving static files.
Hence, it is crucial to
reduce the overhead for the log-writing (producer) application.
We achieve this by assigning  each producer thread  its own ring buffer as can be seen in
Figure~\ref{fig:overview-fips}.
This reduces the complexity to a single-producer, multiple-consumer ring buffer.

Figure~\ref{fig:architecture} shows the
architecture of a single ring with one producer and $n$ consumer applications. A two-phase
approach to enqueue and dequeue elements to the ring is used, similar to the
mechanisms provided by DPDK~\cite{dpdk-zero} and eBPF~\cite{ebpf}. In phase one the producer 
claims memory from the ring buffer, and the application can directly write into
this memory without unnecessary memory copies later. In the second
phase the producer commits this memory to the ring buffer. A consumer later claims the
same memory and can directly operate on the read-only mapped memory. 

The pointers ``Write Head'' and ``Commit Tail'' as shown in the top part of
Figure~\ref{fig:architecture}, are only updated by a single thread. This, therefore, requires no lock or atomic CAS (Compare-and-Swap)  or
FAA (Fetch-and-Add) operation, significantly reducing the overhead for the producer. To ensure
that the ``Commit Tail'' pointer is only moved after the ring buffer cell has been
written to, we use Acquire/Release memory order. Since x86 employs the Total Store
Ordering (TSO) model, these writes become normal \texttt{mov} instructions.
Depending on how the writer should cope with full buffers (as described in
Section~\ref{sec:options}), 
the write head is either moved forward regardless of
whether consumers have released their memory, or it returns \texttt{ENOMEM}.

\begin{figure}
  \centering
  \includegraphics[width=0.5\textwidth]{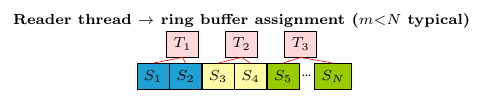}
  \caption{Ring buffer assignment}%
  \label{fig:assignment}
\end{figure}

As discussed previously, the FIPS IPC supports multiple readers reading the same
log messages, with each reader having multiple threads consuming the log messages.
Therefore, each reader has a ``Read'' pointer and a ``Release'' pointer.
A reader can start reading a new message by claiming a new cell, see
\circled{1}, which moves the ``Read'' pointer ahead.
This is done via a CaS operation because multiple reading threads are supported
and to ensure that the ``Read'' pointer does not bypass the ``Commit Tail''. 
CaS operations can be expensive, especially under high contention. To alleviate
this bottleneck, FIPS IPC (a) supports reading in batches (reading multiple
log messages at once) and (b) assigns a consumer thread to multiple ring buffers.
As visualized in Figure~\ref{fig:assignment},
thread $T_2$ is assigned to ring buffer $S_3$ and $S_4$. 
If the CaS operation on ring buffer $S_3$ fails, $T_2$ tries to read a message
from an alternative ring assigned to it, i.e.\ $S_4$ (same color in
Figure~\ref{fig:assignment}). If work stealing is
enabled and none of its assigned rings contain any log messages,
the consumer tries to steal work from another ring buffer
that belongs to a different thread (different color in the figure).

Analyzing string log messages is often done via callbacks (on a match) in the case
of Hyperscan~\cite{hyperscan}, or completely in an asynchronous manner. Using asynchronous
I/O also requires the ring buffer to handle out-of-order releases. When returning a
cell in-order, the release pointer is incremented directly without accessing the
``Release Queue''. This is shown in \circled{2} for consumer $n$. 
Figure~\ref{fig:architecture} shows the
state after consumer 1 has returned cell 6 before cell 5. Cell 6,
therefore, was not returned directly but the cell index is appended to an ordered ``Release
Queue''. Next the consumer returned cell 5, it was returned directly without
accessing the ``Release Queue'' protected by a mutex and moving the release
pointer to index 6. This situation is shown in Figure~\ref{fig:architecture}. If consumer 1 now returns cell
7, a direct return is not possible because the release pointer still points
to cell 6. Before adding a cell index to the ``Release Queue'', it 
checks the front of the queue (see \circled{3}) to see if there are cells that can be returned in-order. 
Consumer 1, therefore, returns the cell 6 and 7 (see \circled{4}) and moves the
``Release'' pointer to cell 8.

By returning cell 6 together with cell 7, in-order returns do not need to access
the mutex-protected ``Release Queue''. This minimizes synchronization overhead for all in order
returns, at the cost of a slightly delayed release pointer update.

\subsection{FIPS IPC API for Writers}\label{subsec:Writer}

\begin{lstlisting}[
  language=C,
  basicstyle=\footnotesize\ttfamily,
  breaklines=true,
  frame=lines,
  numbers=left,
caption={Example usage of the FIPS IPC for an application writing logs},
label={lst:fips-writer},
  framesep=2mm,
  xleftmargin=1.8em
]
% \begin{minted}[fontsize=\footnotesize,breaklines,frame=lines,linenos,
%  framesep=2mm,xleftmargin=1.8em]{c}
fips_buf_arg_t arg = { 
  .role = FIPS_WRITER,
  };
// Needs to be called once at application startup
  fips_buf_init(&arg, "/path/to/config");
// Needs to be called once per thread
  fips_buf_thread_arg_t fips_buf_thread_arg = {0};
  fips_buf_attach(&fips_buf_thread_arg);
// Request a buffer to write log message to
struct iovec iov[NUM_LOGS_TO_WRITE];
fips_get_write_buffer(&fips_buf_thread_arg, iov, NUM_LOGS_TO_WRITE);
// Write log messages to the buffers
// ...
snprinf(iov[i].iov_base, iov[i].iov_len, "Log Message %d", i);
// ...
// Commit the written log messages
fips_buf_write(&fips_buf_thread_arg, NUM_LOGS_TO_WRITE);
\end{lstlisting}

Listing~\ref{lst:fips-writer} shows example usage of the new IPC for an
application, writing logs. First the application must call \texttt{fips\_buf\_init}
to set up the shared memory and ring buffers. It  does not matter whether the
application writing logs or reading logs calls init first. All options can be
configured in the argument struct or via a configuration file. The only
mandatory argument in the struct is the role of the application, either
\texttt{FIPS\_WRITER} or \texttt{FIPS\_READER}. All other options can be set via
the configuration file, making it possible to have one configuration file for
both readers and writers. These options include the size of each ring buffer,
the number of maximum writer threads and readers.

Each thread that wants to write logs must call
\texttt{fips\_buf\_attach} to attach its ring buffer to the thread. 
To write logs the application then requests an arbitrary amount of message buffers to
write logs to using \texttt{fips\_get\_write\_buffer}. After writing the log
messages to the buffers the application must call \texttt{fips\_buf\_write} to
commit the written log messages.

Existing applications can also use the syslog API (\texttt{openlog} \& \texttt{syslog}) to
write logs via the new FIPS IPC\@. This can be done by preloading a shared library
that overrides the syslog functions and uses the new IPC instead. 
This method is called ``FIPS Syslog''.

\subsection{FIPS IPC API for Readers}\label{subsec:Reader}

\begin{lstlisting}[
  language=C,
  basicstyle=\footnotesize\ttfamily,
  breaklines=true,
  frame=lines,
  numbers=left,
  framesep=2mm,
  xleftmargin=1.8em,
  caption={Example usage of the FIPS IPC for an application reading logs},
  label={lst:fips-reader}
]
% \begin{minted}[fontsize=\footnotesize,breaklines,frame=lines,linenos,
%  framesep=2mm,xleftmargin=1.8em]{c}
fips_buf_arg_t arg = { 
  .role = FIPS_READER,
  };
// Needs to be called once at application startup
fips_buf_init(&arg, "/path/to/config");
// Needs to be called once per thread
fips_buf_thread_arg_t fips_buf_thread_arg = {0};
fips_buf_attach(&fips_buf_thread_arg);
// Request a buffer to read log message from
struct iovec iov[NUM_LOGS_TO_READ];
int tag = 0; // tag to identify the buffers for return
int flags = 0; // blocking read & no workload stealing
fips_buf_read(&fips_buf_thread_arg, iov, NUM_LOGS_TO_READ, &tag, flags);
// Process the read log messages
//process_log_message(iov[i].iov_base, iov[i].iov_len);
// ...
// Commit the read log messages
fips_buf_return(&fips_buf_thread_arg, tag);
\end{lstlisting}

For reading, the initialization process is similar as shown in Listing~\ref{lst:fips-reader}.
With \texttt{fips\_buf\_read}, an arbitrary amount of log message buffers can be
requested. After processing these read log messages the application must call
\texttt{fips\_buf\_return} to return the log message buffers.
However, while the writer must commit log buffers in the same order as requested, the
reader can return the log buffers in any order. This allows the reader to
process log messages in parallel and in an asynchronous manner. For this
purpose,
each read request also returns a tag that identifies the requested buffers. This
tag must be provided when returning the buffers. Internally if a reader returns
a buffer out of order the buffer index is stored in a sorted list until all previous
buffers have been returned. To reduce the length of the list we always aggregate
contiguous returned buffers into a single entry in the list. While the
ring buffers themselves are lock-free, the sorted list is protected by a mutex,
because multiple reading threads could be returning messages out of order from
the same ring buffer. 

\subsection{Options for Coping with Full Ring Buffers}\label{sec:options}
In addition to the configuration options mentioned above, the user can also
configure the behavior of the writer if the ring buffer is full.

Three options to handle a full ring buffer are supported:
\begin{itemize}
  \item {\bf Drop}: The writer receives an \texttt{ENOMEM} error when requesting buffer
    space and thus most likely drops the message.
  \item {\bf Overwrite:} The writer will overwrite the oldest log message in the ring
    buffer and move the read pointer forward and accordingly update the out of
    order list.
  \item {\bf Overwrite\_slow\_readers:} The same as above, but if the fastest reader
    would be overwritten, the writer receives an \texttt{ENOMEM} error.
\end{itemize}

The overwrite\_ slow\_readers option allows the fastest reader to at least read
all the log messages in the ring buffer, while still allowing the writer to
proceed if a reader is too slow. If an application requires that all handled client
requests logged, it still can drop the request if the
\texttt{fips\_get\_write\_buffer} returns with no buffer available.

\section{Evaluation}\label{sec:evaluation}

\begin{figure*}[!t]
  \centering
  \includegraphics[width=1\textwidth]{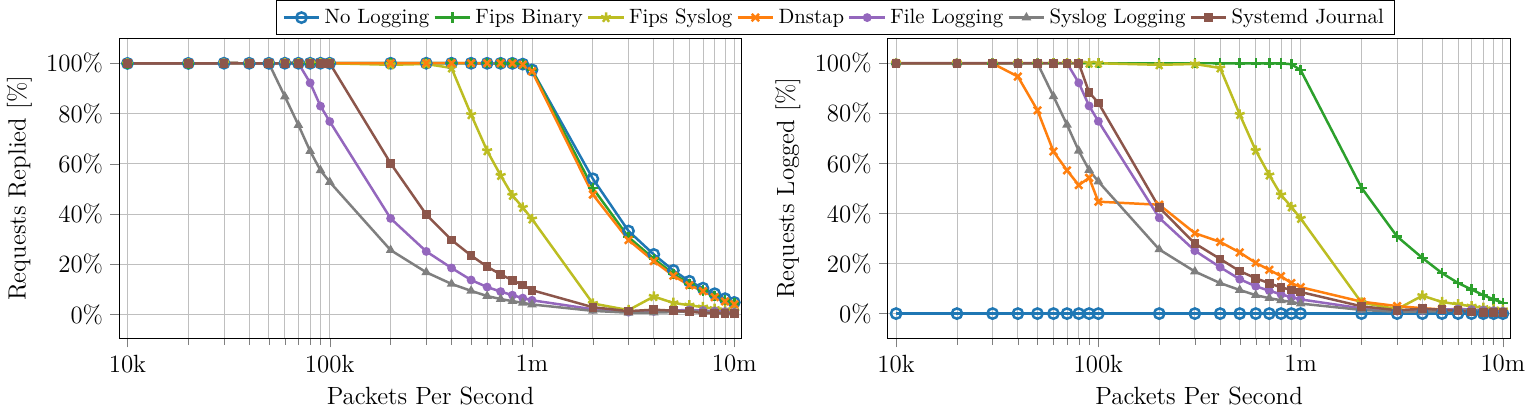}
  \caption{Logging Performance Comparison of BIND9 without Banning Clients}%
  \label{fig:static}
\end{figure*}

The evaluation section is structured as follows.
\begin{enumerate}[label=(\Alph*)]
  \item FIPS IPC is compared against other logging frameworks used by the
    state-of-the-art DNS server BIND9.
  \item The effect of FIPS IPC on client banning is evaluated.
  \item FIPS IPC is also analyzed with the state-of-the-art DNS server BIND9.
\end{enumerate}

All measurements are done on servers with an Intel(R) Xeon(R) Silver 4314 Chip, with
128 GB RAM and ConnectX-6 Dx Cards connected via 100 GBit Ethernet.

For the following experiments FIPS IPC uses a fixed maximal log message size of
$254$ bytes and a total capacity of $2^{17}$ log messages. BIND9 is configured
 with one worker per CPU core, resulting in 16 ring buffers with one per BIND9 worker.
 In our setup with 16 producer, one consumer thread was fast enough to evaluate all
 written log messages. However, we configured FIPS IPC with a maximum reader
 count of 2 and a maximal reader thread count of 16 with each thread having a 
 release queue of 1024 entries. Since only one consumer thread ever attached to the new
 IPC it reads the log messages of all buffers.


The figures present the median of 3 measurements. The source code of FIPS IPC,
its integration into BIND9 and the measurement artifacts are publicly
available\footnote{https://anonymous.4open.science/r/FIPS-IPC-E03F}.

\subsection{Comparison to other Logging Frameworks}

To evaluate FIPS IPC, it was integrated into BIND9 twice. The first integration is
executed before \texttt{isc\_log\_doit} would have been called. It logs data
directly in a binary format without converting elements like IP addresses,
timestamps and query information to strings. This reduces the overhead of
first converting data into strings and later matching it by regular expressions to 
convert it back. This version is labeled ``FIPS Binary'' in Figure~\ref{fig:static}.
The second integration was achieved without any code changes
to BIND9 and without the need to recompile BIND9. It is included by preloading
a shared object, replacing libc's \texttt{openlog} and \texttt{syslog} functions
with those of FIPS IPC (labeled ``FIPS Syslog'').

Figure~\ref{fig:static} shows the percentage of answered DNS
queries and the percentage of logged requests versus the queries per second for different logging frameworks.
We compare the two FIPS versions against all available DNS
  logging options: No Logging, DNSTAP, File Logging, Syslog, and Systemd Journald.
The DNS queries were generated by Trex~\cite{trex} from 255 different IP
addresses. The queries requested DNS records of a small zone file with 10 entries. The workload varies from 10,000 up to
10 million packets per second.

The
figure demonstrates that FIPS 
IPC incurs almost no performance loss compared to no logging. While DNSTAP
logging shows similar performance, the number of logged messages is significantly
lower, as can be seen on the right side of the figure. FIPS Syslog shows the advantages
of the FIPS IPC even when using the syslog API and, in the case of BIND9, being
serialized by a mutex. For all other logging frameworks, the performance is
identical to those presented in Figure~\ref{fig:static-ipc-intro}. 
The number of logged requests is also identical to the number of answered
requests for all logging frameworks except for DNSTAP and no logging. This
highlights the performance benefit of FIPS IPC compared to other logging frameworks.

\subsection{Impact on the IPS}\label{sec:impact-ips}


\begin{figure}
  \centering
  \includegraphics[width=.5\textwidth]{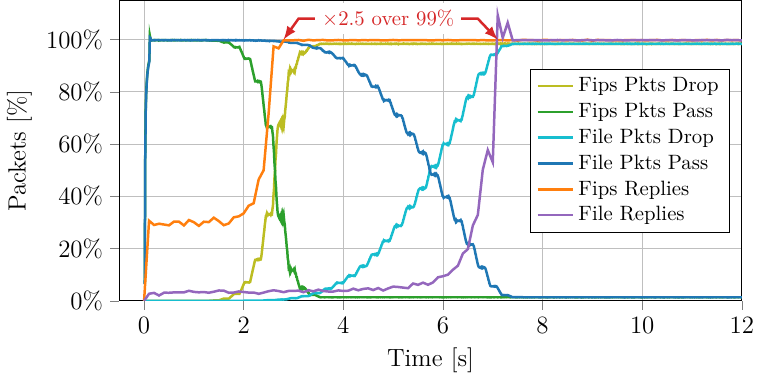}
  \caption{IPC Impact on Ban Speed of simple-Fail2Ban}
  \label{fig:block-speed}
\end{figure}

\begin{figure*}[t]
  \centering
  \includegraphics[width=1\textwidth]{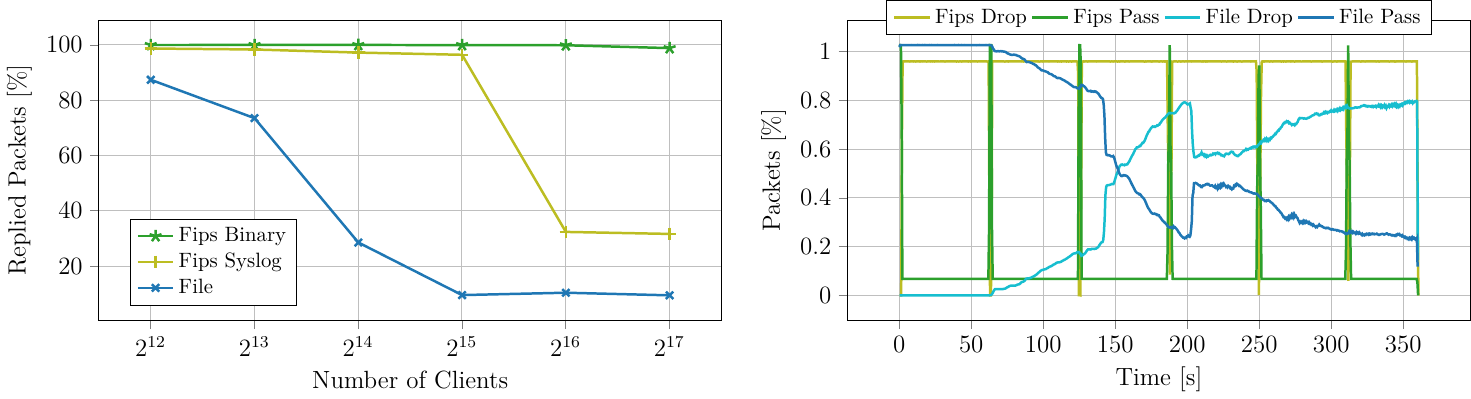}
  \caption{Percentage of DNS Replies (left) and Percentage
    of Packets dropped and passed by simple-Fail2Ban (right)}
  \label{fig:qos}
\end{figure*}

To analyze the impact of the FIPS IPC, we need to eliminate the performance
bottleneck Fail2Ban. We therefore have written a fast test IPS called simple-Fail2Ban. It
mimics the main features of Fail2Ban. It is written in C instead of Python.
It uses Hyperscan~\cite{hyperscan} for fast multi pattern matching instead of
Python's regex engine. Instead of storing clients in a SQLite database it uses
multiple hash tables to store clients and ban times. For all further evaluations
our simple-Fail2Ban implementation is used instead of Fail2Ban. For
the evaluation with IPS simple-Fail2Ban we used file logging as base line, as
its the most prevelent logging option.
In Figure~\ref{fig:block-speed}
we evaluate how the IPC influences the time an IPS needs to ban all DoS IP
addresses. We used the same ban limit of 3 and ban duration of 60 seconds as in
Section~\ref{sec:f2b-performance}.
To stress the new IPC 3,000,000 packets per seconds from a $/16$ subnet were
generated. Additionally, a baseload of 50,000 packets per seconds from a whitelisted
$/24$ subnet was send.
Figure \ref{fig:block-speed} 
shows, for file logging and FIPS IPC, the percentage of packets dropped and passed by the IPS.
The Figure \ref{fig:block-speed}
also depicts how many of the 50,000 requests of the whitelisted clients are
replied by BIND9. The response rate of FIPS IPC to valid
clients is $2.5\times$ faster over $99\%$ than with file logging. While FIPS
reaches 99.5\% at the 2.8s mark, file logging only reaches this value after 7.1s.
The figure also shows that with file logging, the average response rate is
only 3.2\% before any attackers are banned (see ``File Replies''). In contrast, FIPS IPC
achieves an
average response rate of 29.95\% before any attackers are banned (see ``FIPS
Replies'').

Figure~\ref{fig:qos}
shows the influence of the new IPC on the whole system. The left plot shows the
response rate of the whitelisted clients of BIND9 with varying number of
spoofed attack addresses. The
whitelisted clients send $50\,000$ requests per second, while the attacker sends
 a total of $1\,000\,000$ requests per second.
While
Fail2Ban struggled with 1024 clients sending $100\,000$ requests (see Section~\ref{sec:limitations}),  the
FIPS IPC with simple-Fail2Ban easily handles $2^{16}$ clients sending 1 million
requests per second.
The plot also shows the performance of FIPS Syslog on the whole system. This shows
the overhead of the lock in BIND9 \texttt{isc\_log\_doit} function and the
conversion to and from strings. While FIPS Syslog cannot handle $2^{16}$ DoS
clients it shows the advantage of an IPC bypassing the kernel compared to file
logging.
After $2^{14}$ clients the system with file logging answers less than 20\% of the valid
requests. Furthermore, the right side of Figure~\ref{fig:qos} shows that file
logging experiences the same degradation as shown in Figure~\ref{fig:f2b-baseline} with
increasing ban cycles. In contrast FIPS IPC is able to keep up with the log
volume. 


In summary, our experiments show that, once a high-performance IPS which uses
alternatives such as eBPF for packet filtering and Hyperscan for pattern matching is in place, the logging subsystem becomes the dominant
bottleneck which can be eliminated by the FIPS IPC.

\section{Discussion}

In this work, we showed the problems of existing logging frameworks for IPS
systems using BIND9 as an example. Since the FIPS IPC can be used transparently
by any
application supporting the syslog API or by directly using the FIPS API, it can
also be used by other applications. However, the performance impact might vary
depending on the workload generated by each request. Applications with a high
workload per request might see less of a performance impact. However, in our
analysis the high-performance web server nginx also spend 50\% of CPU time
writing query logs when serving static files. This shows the potential of the new
logging IPC. 

io\_uring, the asynchronous file I/O API of Linux,
also relies on ring buffers.
While io\_uring has shown a significant performance improvements for file
  I/O~\cite{io-uring-dbms}, it still relies on syscalls
  (\texttt{io\_uring\_enter}) to submit and
  complete I/O operations~\cite{uring}, which increases overhead compared to
  shared-memory-based approaches. Furthermore, io\_uring has been disabled by default for
  multiple popular projects like Docker, Android and ChromeOS because of
  security concerns~\cite{google-uring,docker-uring,ringguard}.
  While shared memory solutions might also introduce security concerns by
  making reading application transitionally vulnerable from writing
  applications,  the scope is limited to these applications and does not apply
  to the Linux kernel.

  Our evaluation has shown the benefit of the FIPS IPC.
In cooperation with an IPS, 
  BIND9's ability to cope with DoS attacks has improved significantly. The
  integration into BIND9 is just 179 lines of code\footnote{measured with git
  diff}, showing how easily a new logging IPC can be integrated into existing software.
  Furthermore, the ability to be integrated via the widely used Syslog API
  allows effortless integration into existing software.

\section{Future Work}

FIPS IPC has increased the answer rate performance while under attack by 
 6$\times$ compared to file logging. However, the results presented in this work
 did not go beyond
3,000,000 requests per second. Current network interfaces already support 400 Gbps,
and 800 Gbps is on the horizon, making it necessary to further investigate 
IPS performance at higher network speeds. In future work, we want to analyze the
impact of FIPS IPC on TCP applications like nginx. This is currently not included
because Trex, the used load generator, has trouble generating stateful TCP traffic,
when packets are dropped by an IPS.

\section{Conclusion}
In this work, we examined the components of a host-based IPS pipeline and
showed that, once high-performance alternatives such as eBPF for packet
filtering and Hyperscan~\cite{hyperscan} for pattern matching are in place, the logging
subsystem becomes the dominant bottleneck. Using Fail2Ban together with BIND9
as a case study, we demonstrated that an attacker can defeat widely deployed
IPS setups with as little as 65~Mbps of DNS traffic, simply by overwhelming
the file-, syslog-, or journald-based log path.

To address this gap, we introduce FIPS IPC, an IPC  tailored to high-performance
 application logging. FIPS IPC bypasses the kernel by using per-thread lock-free
shared-memory ring buffers, supports multiple independent consumers reading
the same stream at their own pace, and minimizes memory copies.
A drop-in replacement for the syslog API allows existing applications to
benefit from FIPS without source-code changes. The simple native API exposes
the full functionality to applications, further allowing them to
reduce overhead by logging data in a binary format.

Our evaluation with BIND9 showed that the FIPS IPC introduces almost no
overhead compared to fully disabled logging and consistently logs more
requests than any other evaluated framework. When combined with an IPS, FIPS
enables malicious clients to be banned $2.5\times$ faster than with file
logging and sustains workloads of $2^{16}$ attacking clients at one million
requests per second, well beyond the regime in which conventional logging
backends collapse. These results indicate that the logging IPC, rather than
the detection logic, deserves closer attention as link speeds continue to
grow toward 400 and 800~Gbps.

%
\bibliographystyle{IEEEtran} 
\bibliography{fips}
\end{document}